\documentclass[prl,aps,reprint,footinbib,longbibliography,preprintnumbers,superscriptaddress]{revtex4-1}
\usepackage{graphicx}
\usepackage{amssymb}
\usepackage{amsmath}
\usepackage{times}
\usepackage{bm}% bold math
\begin{document}

\title{Low Dimensional Dynamics of Globally Coupled Complex Riccati Equations: \\
Exact Firing-rate Equations for
Spiking Neurons with Clustered Substructure}

 \author{Diego Paz\'o}
 \affiliation{Instituto de F\'{\i}sica de Cantabria (IFCA),
 Universidad de Cantabria-CSIC, 39005 Santander, Spain}

 \author{Rok Cestnik}
 \affiliation{Centre for Mathematical Science, Lund University, 22100 Lund, Sweden}

\date{\today}

\begin{abstract}
We report on an exact
theory for ensembles of globally coupled, heterogeneous
complex Riccati equations. A drastic dimensionality reduction
to a few ordinary differential equations is achieved for Lorentzian heterogeneity.
By applying this technique, we obtain low-dimensional firing-rate
equations for populations of spiking neurons with a clustered substructure.
\end{abstract}

  \maketitle

\paragraph{Introduction.---}
Dimensionality reduction methods are of paramount importance in physics.
Keeping only the essential degrees of
freedom improves our insight into the system behavior,
and increases our chances of controlling it.
Standard dimensionality reduction techniques are approximate.
Remarkably, years ago, Ott and Antonsen discovered an ansatz which permitted
 an exact dimensionality reduction to a few ordinary differential equations (ODEs)
 for populations of heterogeneous phase oscillators \cite{OA08}.
 This method was profusely applied to Kuramoto-like models,
 and pulse-coupled oscillators \cite{LBS13,PM14}, greatly
 improving our understanding of them \cite{BGLM}.
Subsequently,
a close relative of the Ott-Antonsen ansatz, the so-called Lorentzian ansatz,
made it possible to derive low-dimensional closed equations for
ensembles of spiking neurons of
quadratic integrate-and-fire (QIF)
type \cite{MPR15}.
The Lorentzian ansatz was key for deriving exact firing-rate equations (FREs)
\cite{MPR15}, i.e.~for establishing
a theoretical link between ensembles of individual neurons and mesoscopic
firing-rate (or neural-mass) dynamics.
In subsequent years many variants and extensions of
this methodology appeared,
see e.g.~\cite{PM16,RP16,DRM17,DT18,pietras19,MP20,GSK20,goldobin21,chen22,pietras23,clusella24}.
This framework, known as next-generation neural-mass models \cite{CB19,Coombes},
was also used in
practical and clinical studies in neuroscience \cite{rabuffo21,gerster22}.

Exact dimensionality reduction appeared to be
limited to ensembles of one-dimensional units (phase models or QIF neurons).
However, a generalization of the Ott-Antonsen
ansatz for orientable agents on a D-dimensional sphere recently appeared
\cite{chandra19}.
And more recently, one of us found that,
like in certain arrays of phase oscillators \cite{WS93},
arrays of {\em identical} complex-valued Riccati equations are quasi-integrable \cite{rok24}.
Interestingly, the complex Riccati equation, a planar nonlinear system,
appears at an intermediate step in the derivation of the FREs mentioned
above, as well as in calculations with the Ott-Antonsen ansatz. Therefore,
developing a specific dimensionality reduction scheme for ensembles of heterogeneous Riccati equations widens the scope of solvable ensembles
in a non-trivial way.

In this Letter we present
a
dimensionality reduction
scheme
for ensembles of heterogeneous Riccati equations.
The power of this method is exemplified by obtaining exact
low-dimensional FREs for ensembles of QIF neurons with a
clustered substructure.
Numerical simulations confirm the validity of our approach.

\paragraph{Complex-valued Riccati ODEs.---}
We study ensembles of heterogeneous complex-valued Riccati ODEs:
\begin{equation}
 {\dot z}_j=a_j z_j^2+b_j z_j + c_j  \, ,
 \label{ricatti}
\end{equation}
where $z_j(t)\in \mathbb{C}$, and index $j$ runs from $1$ to
the population size $N\gg1$.
Equation~\eqref{ricatti} contains three complex coefficients: $a_j$, $b_j$, and $c_j$,
which depend on the complex mean field
\begin{equation}
Z(t)=\frac1N\sum_{j=1}^N z_j \, .
\label{Z}
 \end{equation}
(Explicit dependence on time is also possible.)
The coefficients $(a_j,b_j,c_j)$ vary across units, as indicated by the
index $j$. Heterogeneity is introduced via (scalar or vector) parameters
${\bm\eta}_j$, which are drawn from a continuous probability density
function (PDF) $G({\bm\eta})$. Hence: $a_j=a({\bm \eta_j},Z,t)$,
likewise for $b_j$ and $c_j$.

\paragraph{Ansatz for the thermodynamic limit.---}
Our theoretical analysis adopts the thermodynamic limit ($N\to\infty$).
We define a conditional density $\rho(z,\bar z|{\bm\eta};t)$,
which satisfies the continuity equation (expressed with complex variables):
\begin{equation}
\partial_t \rho +  \partial_z(\rho \dot z) + \partial_{\bar z}(\rho \dot{\bar z})=0 \, ,
\label{cont}
\end{equation}
here $\bar z$ stands for the complex conjugate of $z$.
We here present the following ansatz as an exact solution of the continuity equation:
\begin{equation}
\rho (z,\bar z|{\bm\eta};t)=\frac{\alpha({\bm\eta},t)^2}
{\pi  \left(\left|z-q({\bm\eta},t)\right|^2+\alpha({\bm\eta},t)^2 \right)^2} \, .
\label{ansatz}
\end{equation}
This density is bell-shaped, with its center located
at $q\in\mathbb{C}$, and
radial width
$\langle |z-q|\rangle = (\pi/2)\alpha \ge0$.

The form of \eqref{ansatz}
was suggested by the stereographic projection of a uniform
density on the surface of the unit 3-D sphere onto the complex plane,
with an offset $q$ and
a width $\alpha$ as free parameters. Originally, the motivation
for this stems from
the mapping between the dynamics of
an orientable agent on a 3-D sphere
and the Riccati dynamics with real coefficients,  see \cite{pazo24} for illustration.

The functional form of the density \eqref{ansatz} is invariant under
time evolution, with $q$ and $\alpha$ obeying:
\begin{subequations}
 \begin{eqnarray}
\partial_t q({\bm\eta},t)&=& a q^2+b q + c-\bar a \alpha^2 \, , \label{riccatia} \\
\partial_t \alpha({\bm\eta},t)&=& \left[aq+\bar a \bar q +(b +\bar b)/2\right]\alpha \,.
\label{ricattib}
\end{eqnarray}
\label{quadratic}
\end{subequations}
Closing this system of equations requires determining the mean field.
For a given $\bm\eta$ value, the average is located at $q({\bm\eta},t)$,
and the continuous counterpart of \eqref{Z} is obtained by integrating
$q$ over $\bm\eta$:
\begin{equation}
Z(t)= \int G({\bm\eta}) \, q({\bm\eta},t) \, d{\bm\eta}
\label{barz}
\end{equation}
Equations \eqref{quadratic} and \eqref{barz} define a
closed system of integro-differential equations,
which exactly governs the dynamics in the subspace
determined by our density ansatz  \eqref{ansatz}.

\paragraph{Reduction to ODEs.---}
Further reduction to ODEs is possible in certain cases only.
We focus hereafter on one of these cases.
From now on,
 only the real part of $c_j$ is heterogeneous:
\begin{subequations}
\begin{eqnarray}
a_j&=& a(Z,t) , \\
b_j&=& b(Z,t) ,\\
c_j&=&\eta_j+i\, \Gamma + f(Z,t) \, ,
\end{eqnarray}
\end{subequations}
where
$\eta_j$ and $\Gamma$  are real, and
$f=f_r+if_i$ is a complex function.
Hence, in the $N\to\infty$ limit,
the mean field
is obtained by integrating over
the  only source of heterogeneity:
\begin{equation}
Z(t)= \int\limits_{-\infty}^\infty g(\eta) \, q(\eta,t)\,  d\eta \, ,
\label{int}
\end{equation}
where $g(\eta)$ is the PDF of $\eta$.

A particularly sharp reduction of dimensionality is
achieved for a Lorentzian distribution of the
heterogeneity, $\eta_j\sim L(\eta_j|\eta_0,\delta$):
\begin{equation}
g(\eta)=L(\eta|\eta_0,\delta)=
\frac{\delta/\pi}{(\eta-\eta_0)^2+
\delta^2} .
\label{lor}
\end{equation}
Here $\eta_0$ is the center of the distribution, and $\delta>0$ denotes its half-width at half-maximum (HWHM).
We can evaluate the integral in Eq.~\eqref{int} using the residue theorem. 
By closing the integration path with an arc at complex infinity, the integral takes the value 
of $q(\eta_p,t)$, where $\eta_p$ is the pole of $g(\eta)=(2\pi i)^{-1}[(\eta-\eta_0-i\delta)^{-1}-(\eta-\eta_0+i\delta)^{-1}]$ 
located inside the integration contour --- provided $q$ is analytic inside.
The suitable pole turns out to be \footnote{If the sign of $\Gamma+f_i-{b_r  b_i}/(2a)$ changes under the time evolution, then our approach breaks down (one cannot simply flip the sign of the $\pm$). Fortunately, that does not occur in many situations.
See Supplemental Material at xxx for details.}
\begin{equation}
\eta_p=\eta_0\pm i\delta, \, \, \, \, \mbox{with \,  $\pm= \mathrm{sgn}\left(\Gamma+f_i-\frac{b_r  b_i}{2a}\right)$ ,}
\end{equation}
with $b=b_r+i\,  b_i$, and assuming real $a>0$.
In this way the mean field in Eq.~\eqref{int} becomes:
\begin{equation}
Z(t)=q(\eta_p,t) .
\end{equation}
Evaluating Eqs.~\eqref{quadratic} at $\eta_p$ dramatically
reduces the dimensionality of the problem.
For closing this system of ODEs we also
have to evaluate the variable $\bar q$ at the pole $\eta_p$.
Noting that $\bar q(\eta,t)$ is governed by Eq.~\eqref{riccatia} under complex conjugation, we end up with
a system of
three
complex-valued ODEs for variables
$\{Z,A,Q\}=\{q,\alpha,\bar q\}_{\eta=\eta_p}$:
\begin{subequations}
\begin{eqnarray}
\dot Z&=& a Z^2+b Z+ \eta_p +i \Gamma + f -  a A^2 \, ,  \label{Zdot}\\
 \dot A&=& [a (Z + Q) + (b+\bar b)/2] A \label{A} \, ,\\
 \dot Q &=& a Q^2 + \bar b Q + \eta_p - i \Gamma + \bar f - a A^2 \, .
\end{eqnarray}
\label{3odes}
\end{subequations}
This system
describes the dynamics
of an infinite ensemble of Riccati equations inside the manifold
defined by densities of the form in Eq.~\eqref{ansatz}.
Next, we showcase the reduced Eqs.~\eqref{3odes} on
a physically meaningful example.

\paragraph{Firing-rate equations for clustered neurons.---}
Firing-rate
models constitute useful representations
of the activity of ensembles of neurons
in terms of a few collective variables \cite{ET10,Coombes}.
A decade ago, an
exact reduction to a system of two ODEs was achieved for ensembles
of globally coupled QIF neurons \cite{MPR15}.

Here we go one step further, and consider
a population consisting of
$N$ subpopulations of internally coupled QIF neurons.
The microscopic equations are:
\begin{eqnarray}
{\dot V}_{i,j}&=&V_{i,j}^2+\eta_{i,j} + \phi(r_j) + f(R,t) \label{v}\\
r_j&=&\frac{1}M \sum_{i=1}^M \sum_{k} \delta^{(\epsilon)}(t-t_{i,j}^k) \label{rj}\\
R&=&\frac{1}N\sum_{j=1}^N r_j
\label{qif}
\end{eqnarray}
 Here $V_{i,j}(t)\in  \mathbb{R}$ is the membrane potential of the $i$-th neuron,
 in the $j$-th subpopulation.
 The $\delta^{(\epsilon)}$
 function in Eq.~\eqref{rj} represents the spikes at firing times $t_{i,j}^k$ (when $V_{i,j}$
reaches $\infty$ and  is reset to $-\infty$),
and becomes the Dirac delta after taking $M\to\infty$.
The $r_j$'s are average firing rates of subpopulations, while $R$ is the global firing rate.
The internal coupling in each subpopulation is represented
by the  function $\phi(r_j)$, while the coupling $f$
in Eq.~\eqref{v} is the result of every QIF neuron interacting
with all the other ones at the global scale.

In the $M\to \infty$ limit,
for Lorentzian distributed internal currents
$\eta_{i,j} \sim L(\eta_{i,j}|\eta_j,\Delta)$,
applying the method in \cite{MPR15} yields
an ensemble of globally coupled FREs.
The state of each node is given by the firing rate $r_j$ and
 the membrane voltage $v_j= M^{-1}$ $\sum_i V_{i,j}$.
Their evolution is governed by ODEs:
\begin{subequations}
 \begin{eqnarray}
\dot v_j&=&v_j^2 -\pi^2 r_j^2  + \eta_j  + \phi(r_j) + f(R,t) \\
\dot r_j&=& 2   v_j r_j+ \frac{\Delta}\pi
\end{eqnarray}
\label{nmm}
\end{subequations}
Recall that parameters $\eta_j$ and $\Delta>0$ are the center and the HWHM of the Lorentzian distribution of internal currents
$\eta_{i,j}$
in the $j$-th subpopulation of QIF neurons \cite{MPR15}.
More generally, $\Delta$ can represent the sum of the HWHM and
the amplitude of Cauchy noise \cite{pietras23,clusella24}.
The main variables and parameters are summarized in Table~\ref{table}.

\begin{table}
\centering % used for centering table
\begin{tabular}{c c    c  c } % centered columns (5 columns)
\hline\hline %inserts double horizontal lines
\phantom & Voltages &  Rates & Internal Parameters \\ %[0.5ex] % inserts table
%heading
\hline \\ [-1ex]% inserts single horizontal line
Neurons &  $V_{i,j}$ &  --- &
$\, \eta_{i,j}\sim L(\eta_{i,j}|\eta_j,\Delta)$ \\  [1ex]
Subpopulations& $v_j$ & $r_j$   & $\, \Delta$ and $\eta_j\sim L(\eta_j|\eta_0,\delta)$ \\[1ex] % [1ex] adds vertical space
Whole population &
$V$
& $R$   & $\delta$ and $\eta_0$
\\[1ex] % [1ex] adds vertical space
\hline\hline
\end{tabular}
%\label{table}
\caption{Variables and internal parameters at each description level for the two examples
of dimensionality reduction in this Letter.
$L(h|h_0,d)$ indicates $h$ is Lorentzian distributed with
center at $h_0$ and HWHM $d$. Coupling parameters depend on the specific example and are not
included in the Table.}
\label{table}
\end{table}

For physical reasons $\phi$ has to be a monotonic function
with $\phi(0)=0$.
For solvability, we choose
the quadratic function $\phi(r)=\kappa \, r^2$.
This is different from the standard linear function, but
can be used to approximate a linear interaction around
a reference firing rate $r_0$: $r\approx r^2/(2 r_0)-r_0/2$; or it
can appear from triplet interactions.
The constant $\kappa$ represents inhibitory coupling
for $\kappa<0$, and excitatory coupling for $\kappa>0$.

For $\kappa<\pi^2$ we can define the
complex variable
\begin{equation}
z_j\equiv v_j+i (\pi^2-\kappa)^{1/2}  r_j ,
\label{transformation}
\end{equation}
such that
Eq.~\eqref{nmm} becomes an ensemble of globally coupled Riccati equations:
\begin{equation}
\dot z_j=z_j^2+\eta_j + i(1-\kappa/\pi^2)^{1/2} \Delta + f( R,t)
\label{nmmc}
\end{equation}
We adopt the standard chemical coupling
$f=J R$, where $J$ is the coupling constant,
and $R=\mathrm{Im}(Z)/(\pi^2-\kappa)^{1/2}$.
In Fig.~1 we show the result of simulating an ensemble of $10^4$
units with Lorentzian distributed $\eta_j$, and
an initial condition $z_j(0)$ drawn randomly from the density
\eqref{ansatz}
with $\eta$-independent parameters: $q(\eta,0)=q_0$ and $\alpha(\eta,0)=\alpha_0$.
The two panels in Fig.~1 differ
only in
the value of $\alpha_0$: $0.5$ and $2$ for (a) and (b), respectively.
It may be appreciated that the average $Z$ behaves
in very good agreement with the prediction of the
corresponding 6-dimensional
model \eqref{3odes}. The conformity between the simulation
and the theory is remarkable, especially considering that the two different initial conditions lead to distinct
attractors.

\begin{figure}
	\includegraphics[width=0.9\linewidth]{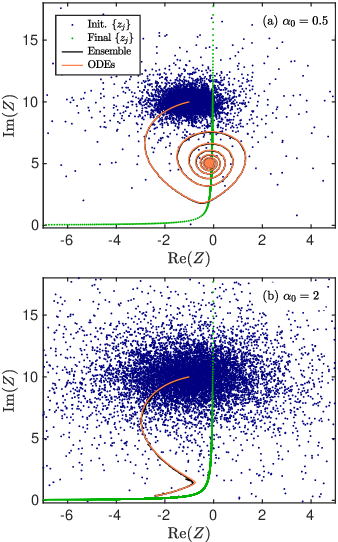}
	\caption{Phase portraits of the mean field $Z$ for two different initial conditions.
	The black line is the trajectory of $Z$ obtained by simulating an ensemble of $10^4$ units, while the orange line
	shows the prediction from the reduced model of 3 complex-valued ODEs, Eq.~\eqref{3odes}.
	System parameters are: $\kappa=\pi^2/2$, $J=16$, $\eta_0=-8$, $\Delta=\delta=1$.
	Initial conditions were chosen according to the ansatz \eqref{ansatz} with $q_0=-1+10i$, and the only difference between the two panels is the width of the initial distribution $\alpha_0$, set to 0.5 in (a) and 2 in (b).
	The initial and the final state of the ensemble of units $\{z_j\}_{j=1,\ldots,N}$ are
	depicted with blue and green dots, respectively.
	Animations of the figures are available in the Supplemental Material.
	}
\label{pd}
\end{figure}

For the system \eqref{nmmc}, $A$ always approaches zero
asymptotically. This follows from the coincidence (up to a factor 6)
between the real part of the coefficient accompanying
$A$ in Eq.~\eqref{A} and
the trace of the Jacobian of the system,
$6[a \mathrm{Re}(Z+Q)+b_r]$, which
determines the volume contraction rate and is negative on average \cite{PP_LEs}
(unless the system is conservative).
This means that the asymptotic
behaviors of the system
are well described by Eq.~\eqref{Zdot} alone, setting
$A=0$.
One could have derived that equation assuming from the outset
that there is a (smooth) univocal dependence of $z_j$ on
the value of $\eta_j$ (i.e., $\alpha=0$).
This assumption enables the transformation $z_j(t) \to z(\eta,t)$ and immediately
leads to Eq.~\eqref{riccatia} with $\alpha=0$.

It is instructive to revert the change of variables defined in
\eqref{transformation} and
write down the equation
for the asymptotic dynamics of mean voltage
$V=N^{-1}\sum_j v_j$ and firing rate $R$:
\begin{subequations}
\begin{eqnarray}
\dot V &=&V^2-(\pi^2-\kappa) R^2+ \eta_0 + J R \, \\
\dot R &=&  2 V R  + \frac{\Delta}\pi + \frac{\delta}{(\pi^2-\kappa)^{1/2}} \, .
\end{eqnarray}
\label{mf_nmm}
\end{subequations}
These two ODEs exactly describe the asymptotic global behavior of an ensemble of
QIF neurons with two levels of heterogeneity and
coupling. The microscopic level depends on
parameters $\Delta$ and $\kappa$, while at the mesoscopic
level $\delta$ and $J$ play that role.
The phase diagram of system \eqref{mf_nmm}
is trivial, since the
FREs in \cite{MPR15} are recovered,
after appropriate rescaling of $R$, and defining effective $J$ and $\Delta$
parameters.
In particular, for $\kappa=0$ we consistently recover the FREs
for Lorentzian heterogeneity of currents centered
at $\eta_0$ and HWHM $\Delta+\delta$
\footnote{This is precisely the marginal distribution for
a conditional Lorentzian distribution $L(\eta|\eta_c,\Delta)$
centered at $\eta_c$ with HWHM $\Delta$,
if $\eta_c$ is itself
Lorentzian-distributed with center $\eta_0$
and HWHM $\delta$:
$\int d\eta_c L(\eta|\eta_c,\Delta) L(\eta_c|\eta_0,\delta)
=L(\eta|\eta_0,\delta+\Delta)$.}.

\paragraph{
Global
electrical
coupling.---}
The synapses between neurons can be either chemical or electrical
(gap junctions).
Electrical coupling may trigger
 unsteady periodic dynamics in one single population of QIF neurons without additional ingredients such as delay \cite{pietras19}.
Hence, we modify our ensemble of QIF neurons
considering a global electrical coupling
instead of chemical coupling.
The form of the internal coupling $\phi(r_j)$ is preserved.
The ensemble of FREs turns out to obey \footnote{The microscopic model
of QIF neurons leading to Eq.~\eqref{nmme},
via \cite{pietras19,MP20}, is
available in the Supplemental Material at XXXX.}:
\begin{subequations}
 \begin{eqnarray}
\dot v_j&=&v_j^2 -(\pi^2-\kappa) r_j^2  + \eta_j  + g (V-v_j) \, ,\\
\dot r_j&=& 2   v_j r_j+ \frac{\Delta}\pi -g r_j \, .
\end{eqnarray}
\label{nmme}
\end{subequations}
Now the global coupling actuates
through the mean voltage
$V(t)$, and $g>0$ is the conductance.
A global chemical coupling term $J\,  R$ could also be included, but we opted
not to do so for simplicity.

We apply transformation \eqref{transformation}
and our reduction theory to Eq.~\eqref{nmme}.
If we exclusively focus on
the asymptotic dynamics, we need only
the  complex-valued ODE \eqref{Zdot} with
$A=0$.
Written in terms of $V$ and $R$, it yields exact
macroscopic FREs:
\begin{subequations}
\begin{eqnarray}
\dot V &=&V^2-(\pi^2-\kappa) R^2+ \eta_0 \, ,\\
\dot R &=&  2 V R  + \frac{\Delta}\pi + \frac{\delta}{(\pi^2-\kappa)^{1/2}}-g R \, .
\end{eqnarray}
\label{mf_nmme}
\end{subequations}

We test the validity of these FREs combining
global electrical coupling with inhibitory
internal coupling ($\kappa<0$) in each subpopulation.
Figure 2 shows the comparison between the
oscillatory behavior of the ensemble \eqref{nmme}
and the limit cycle displayed by the
system of ODEs \eqref{mf_nmme}.
Again, we find very good agreement between theory and simulation.

\begin{figure}
	\includegraphics[width=0.9\linewidth]{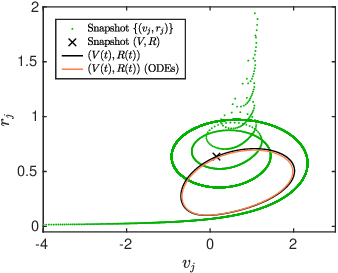}
	\caption{
 	Snapshot of the coordinates of system~\eqref{nmme} with  $N=8000$
	in its asymptotic regime, see green dots.
The location of the mean field is signaled by a $\times$.
	The trajectory of the mean field $(V(t),R(t))$ extracted from the
	direct simulation of Eq.~\eqref{nmme}, and
the limit cycle predicted by Eq.~\eqref{mf_nmme} are depicted by black and
orange solid lines, respectively.
Parameters: $g=2.5$, $\kappa=-\pi^2$, $\eta_0=1$, and $\Delta=\delta=0.5$.
}
\label{fig:elec}
\end{figure}

\paragraph{Conclusions.---} In this work
we have uncovered
an exact
dimensionality reduction scheme
for an infinite set of globally coupled complex-valued Riccati equations.
Our approach begins with an ansatz for the continuity equation,
which defines a lower dimensional manifold.
The application of the residue theorem further reduced the dimensionality
to only three complex-valued ODEs, Eq.~\eqref{3odes}.
The mathematical analysis assumed that if an initial $q(\eta,0)$
possesses an analytic continuation in a particular complex $\eta$ half-plane, it remains analytic under time evolution. 
This might not be completely
justified since $\bar q(\eta,t)$ does not
remain analytic
in the same half-plane, and it drives $q$ via $\alpha$.
This limitation raises the question of whether the
six-dimensional
system \eqref{3odes} is merely an excellent approximation.
In the large-time limit, $A(t)$ almost always approaches $0$,
and the resulting asymptotic two-dimensional system
exactly describes the attractors (in the thermodynamic limit).

We applied our theory to ensembles of
FREs, confirming the validity of our approach.
All simulations showed good agreement with the theory.
The reduced six-dimensional model
consistently predicted the correct asymptotic states across initial conditions,
in spite of the analyticity issue
discussed above.
Moreover, the two-dimensional systems, Eqs.~\eqref{mf_nmm} and \eqref{mf_nmme}
perfectly described the attractors.

It is noteworthy that with this formalism one could consider arbitrarily nested substructures, since at each step the equations can be of the Riccati form given a certain form of coupling, e.g. for $f=J\,R^2$, Eq.~\eqref{mf_nmm} is equivalent to a Riccati equation \footnote{See Supplemental Material at XXX.}.

Our methodology represents a significant breakthrough
in dimensional reductions as it expands the solvable
examples from traditional
one-dimensional units into the realm of planar units.
This opens up many new examples to explore,
with many unit and coupling dynamics one can choose from.
Additionally, the heterogeneity may enter
in several ways (not only in the real part of $c_j$). 
Beyond the variations to the FREs that were studied here, we can also mention the ``complexified'' Kuramoto model~\cite{complexified2023,complexified2024} as a system where
our methodology can be applied.

\begin{acknowledgments}
DP acknowledges support by
Grant No. PID2021-125543NB-I00,
funded by MICIU/AEI/10.13039/501100011033 and by ERDF/EU.
RC acknowledges financial support from the Royal Swedish
Physiographic Society of Lund.
 \end{acknowledgments}

%     \bibliography{bibliografia}

\begin{thebibliography}{33}%
\makeatletter
\providecommand \@ifxundefined [1]{%
 \@ifx{#1\undefined}
}%
\providecommand \@ifnum [1]{%
 \ifnum #1\expandafter \@firstoftwo
 \else \expandafter \@secondoftwo
 \fi
}%
\providecommand \@ifx [1]{%
 \ifx #1\expandafter \@firstoftwo
 \else \expandafter \@secondoftwo
 \fi
}%
\providecommand \natexlab [1]{#1}%
\providecommand \enquote  [1]{``#1''}%
\providecommand \bibnamefont  [1]{#1}%
\providecommand \bibfnamefont [1]{#1}%
\providecommand \citenamefont [1]{#1}%
\providecommand \href@noop [0]{\@secondoftwo}%
\providecommand \href [0]{\begingroup \@sanitize@url \@href}%
\providecommand \@href[1]{\@@startlink{#1}\@@href}%
\providecommand \@@href[1]{\endgroup#1\@@endlink}%
\providecommand \@sanitize@url [0]{\catcode `\\12\catcode `\$12\catcode
  `\&12\catcode `\#12\catcode `\^12\catcode `\_12\catcode `\%12\relax}%
\providecommand \@@startlink[1]{}%
\providecommand \@@endlink[0]{}%
\providecommand \url  [0]{\begingroup\@sanitize@url \@url }%
\providecommand \@url [1]{\endgroup\@href {#1}{\urlprefix }}%
\providecommand \urlprefix  [0]{URL }%
\providecommand \Eprint [0]{\href }%
\providecommand \doibase [0]{http://dx.doi.org/}%
\providecommand \selectlanguage [0]{\@gobble}%
\providecommand \bibinfo  [0]{\@secondoftwo}%
\providecommand \bibfield  [0]{\@secondoftwo}%
\providecommand \translation [1]{[#1]}%
\providecommand \BibitemOpen [0]{}%
\providecommand \bibitemStop [0]{}%
\providecommand \bibitemNoStop [0]{.\EOS\space}%
\providecommand \EOS [0]{\spacefactor3000\relax}%
\providecommand \BibitemShut  [1]{\csname bibitem#1\endcsname}%
\let\auto@bib@innerbib\@empty
%</preamble>
\bibitem [{\citenamefont {Ott}\ and\ \citenamefont {Antonsen}(2008)}]{OA08}%
  \BibitemOpen
  \bibfield  {author} {\bibinfo {author} {\bibfnamefont {E.}~\bibnamefont
  {Ott}}\ and\ \bibinfo {author} {\bibfnamefont {T.~M.}\ \bibnamefont
  {Antonsen}},\ }\bibfield  {title} {\enquote {\bibinfo {title} {Low
  dimensional behavior of large systems of globally coupled oscillators},}\
  }\href {\doibase 10.1063/1.2930766} {\bibfield  {journal} {\bibinfo
  {journal} {Chaos}\ }\textbf {\bibinfo {volume} {18}},\ \bibinfo {eid}
  {037113} (\bibinfo {year} {2008})}\BibitemShut {NoStop}%
\bibitem [{\citenamefont {Luke}\ \emph {et~al.}(2013)\citenamefont {Luke},
  \citenamefont {Barreto},\ and\ \citenamefont {So}}]{LBS13}%
  \BibitemOpen
  \bibfield  {author} {\bibinfo {author} {\bibfnamefont {T.~B.}\ \bibnamefont
  {Luke}}, \bibinfo {author} {\bibfnamefont {E.}~\bibnamefont {Barreto}}, \
  and\ \bibinfo {author} {\bibfnamefont {P.}~\bibnamefont {So}},\ }\bibfield
  {title} {\enquote {\bibinfo {title} {Complete classification of the
  macroscopic behavior of a heterogeneous network of theta neurons},}\
  }\href@noop {} {\bibfield  {journal} {\bibinfo  {journal} {Neural Comput.}\
  }\textbf {\bibinfo {volume} {25}},\ \bibinfo {pages} {3207--3234} (\bibinfo
  {year} {2013})}\BibitemShut {NoStop}%
\bibitem [{\citenamefont {Paz\'o}\ and\ \citenamefont
  {Montbri\'o}(2014)}]{PM14}%
  \BibitemOpen
  \bibfield  {author} {\bibinfo {author} {\bibfnamefont {D.}~\bibnamefont
  {Paz\'o}}\ and\ \bibinfo {author} {\bibfnamefont {E.}~\bibnamefont
  {Montbri\'o}},\ }\bibfield  {title} {\enquote {\bibinfo {title}
  {Low-dimensional dynamics of populations of pulse-coupled oscillators},}\
  }\href {\doibase 10.1103/PhysRevX.4.011009} {\bibfield  {journal} {\bibinfo
  {journal} {Phys. Rev. X}\ }\textbf {\bibinfo {volume} {4}},\ \bibinfo {pages}
  {011009} (\bibinfo {year} {2014})}\BibitemShut {NoStop}%
\bibitem [{\citenamefont {Bick}\ \emph {et~al.}(2020)\citenamefont {Bick},
  \citenamefont {Goodfellow}, \citenamefont {Laing},\ and\ \citenamefont
  {Martens}}]{BGLM}%
  \BibitemOpen
  \bibfield  {author} {\bibinfo {author} {\bibfnamefont {C.}~\bibnamefont
  {Bick}}, \bibinfo {author} {\bibfnamefont {M.}~\bibnamefont {Goodfellow}},
  \bibinfo {author} {\bibfnamefont {C.~R.}\ \bibnamefont {Laing}}, \ and\
  \bibinfo {author} {\bibfnamefont {E.~A.}\ \bibnamefont {Martens}},\
  }\bibfield  {title} {\enquote {\bibinfo {title} {Understanding the dynamics
  of biological and neural oscillator networks through exact mean-field
  reductions: a review},}\ }\href@noop {} {\bibfield  {journal} {\bibinfo
  {journal} {J. Math. Neurosci.}\ }\textbf {\bibinfo {volume} {10}},\ \bibinfo
  {pages} {9} (\bibinfo {year} {2020})}\BibitemShut {NoStop}%
\bibitem [{\citenamefont {Montbri\'o}\ \emph {et~al.}(2015)\citenamefont
  {Montbri\'o}, \citenamefont {Paz\'o},\ and\ \citenamefont {Roxin}}]{MPR15}%
  \BibitemOpen
  \bibfield  {author} {\bibinfo {author} {\bibfnamefont {E.}~\bibnamefont
  {Montbri\'o}}, \bibinfo {author} {\bibfnamefont {D.}~\bibnamefont {Paz\'o}},
  \ and\ \bibinfo {author} {\bibfnamefont {A.}~\bibnamefont {Roxin}},\
  }\bibfield  {title} {\enquote {\bibinfo {title} {Macroscopic description for
  networks of spiking neurons},}\ }\href {\doibase 10.1103/PhysRevX.5.021028}
  {\bibfield  {journal} {\bibinfo  {journal} {Phys. Rev. X}\ }\textbf {\bibinfo
  {volume} {5}},\ \bibinfo {pages} {021028} (\bibinfo {year}
  {2015})}\BibitemShut {NoStop}%
\bibitem [{\citenamefont {Paz\'o}\ and\ \citenamefont
  {Montbri\'o}(2016)}]{PM16}%
  \BibitemOpen
  \bibfield  {author} {\bibinfo {author} {\bibfnamefont {D.}~\bibnamefont
  {Paz\'o}}\ and\ \bibinfo {author} {\bibfnamefont {E.}~\bibnamefont
  {Montbri\'o}},\ }\bibfield  {title} {\enquote {\bibinfo {title} {From
  quasiperiodic partial synchronization to collective chaos in populations of
  inhibitory neurons with delay},}\ }\href {\doibase
  10.1103/PhysRevLett.116.238101} {\bibfield  {journal} {\bibinfo  {journal}
  {Phys. Rev. Lett.}\ }\textbf {\bibinfo {volume} {116}},\ \bibinfo {pages}
  {238101} (\bibinfo {year} {2016})}\BibitemShut {NoStop}%
\bibitem [{\citenamefont {Ratas}\ and\ \citenamefont {Pyragas}(2016)}]{RP16}%
  \BibitemOpen
  \bibfield  {author} {\bibinfo {author} {\bibfnamefont {I.}~\bibnamefont
  {Ratas}}\ and\ \bibinfo {author} {\bibfnamefont {K.}~\bibnamefont
  {Pyragas}},\ }\bibfield  {title} {\enquote {\bibinfo {title} {Macroscopic
  self-oscillations and aging transition in a network of synaptically coupled
  quadratic integrate-and-fire neurons},}\ }\href {\doibase
  10.1103/PhysRevE.94.032215} {\bibfield  {journal} {\bibinfo  {journal} {Phys.
  Rev. E}\ }\textbf {\bibinfo {volume} {94}},\ \bibinfo {pages} {032215}
  (\bibinfo {year} {2016})}\BibitemShut {NoStop}%
\bibitem [{\citenamefont {Devalle}\ \emph {et~al.}(2017)\citenamefont
  {Devalle}, \citenamefont {Roxin},\ and\ \citenamefont {Montbri\'o}}]{DRM17}%
  \BibitemOpen
  \bibfield  {author} {\bibinfo {author} {\bibfnamefont {F.}~\bibnamefont
  {Devalle}}, \bibinfo {author} {\bibfnamefont {A.}~\bibnamefont {Roxin}}, \
  and\ \bibinfo {author} {\bibfnamefont {E.}~\bibnamefont {Montbri\'o}},\
  }\bibfield  {title} {\enquote {\bibinfo {title} {Firing rate equations
  require a spike synchrony mechanism to correctly describe fast oscillations
  in inhibitory networks},}\ }\href {\doibase 10.1371/journal.pcbi.1005881}
  {\bibfield  {journal} {\bibinfo  {journal} {PLoS Computational Biology}\
  }\textbf {\bibinfo {volume} {13}},\ \bibinfo {pages} {1--21} (\bibinfo {year}
  {2017})}\BibitemShut {NoStop}%
\bibitem [{\citenamefont {di~Volo}\ and\ \citenamefont {Torcini}(2018)}]{DT18}%
  \BibitemOpen
  \bibfield  {author} {\bibinfo {author} {\bibfnamefont {M.}~\bibnamefont
  {di~Volo}}\ and\ \bibinfo {author} {\bibfnamefont {A.}~\bibnamefont
  {Torcini}},\ }\bibfield  {title} {\enquote {\bibinfo {title} {Transition from
  asynchronous to oscillatory dynamics in balanced spiking networks with
  instantaneous synapses},}\ }\href {\doibase 10.1103/PhysRevLett.121.128301}
  {\bibfield  {journal} {\bibinfo  {journal} {Phys. Rev. Lett.}\ }\textbf
  {\bibinfo {volume} {121}},\ \bibinfo {pages} {128301} (\bibinfo {year}
  {2018})}\BibitemShut {NoStop}%
\bibitem [{\citenamefont {Pietras}\ \emph {et~al.}(2019)\citenamefont
  {Pietras}, \citenamefont {Devalle}, \citenamefont {Roxin}, \citenamefont
  {Daffertshofer},\ and\ \citenamefont {Montbri\'o}}]{pietras19}%
  \BibitemOpen
  \bibfield  {author} {\bibinfo {author} {\bibfnamefont {B.}~\bibnamefont
  {Pietras}}, \bibinfo {author} {\bibfnamefont {F.}~\bibnamefont {Devalle}},
  \bibinfo {author} {\bibfnamefont {A.}~\bibnamefont {Roxin}}, \bibinfo
  {author} {\bibfnamefont {A.}~\bibnamefont {Daffertshofer}}, \ and\ \bibinfo
  {author} {\bibfnamefont {E.}~\bibnamefont {Montbri\'o}},\ }\bibfield  {title}
  {\enquote {\bibinfo {title} {Exact firing rate model reveals the differential
  effects of chemical versus electrical synapses in spiking networks},}\ }\href
  {\doibase 10.1103/PhysRevE.100.042412} {\bibfield  {journal} {\bibinfo
  {journal} {Phys. Rev. E}\ }\textbf {\bibinfo {volume} {100}},\ \bibinfo
  {pages} {042412} (\bibinfo {year} {2019})}\BibitemShut {NoStop}%
\bibitem [{\citenamefont {Montbri\'o}\ and\ \citenamefont
  {Paz\'o}(2020)}]{MP20}%
  \BibitemOpen
  \bibfield  {author} {\bibinfo {author} {\bibfnamefont {E.}~\bibnamefont
  {Montbri\'o}}\ and\ \bibinfo {author} {\bibfnamefont {D.}~\bibnamefont
  {Paz\'o}},\ }\bibfield  {title} {\enquote {\bibinfo {title} {Exact mean-field
  theory explains the dual role of electrical synapses in collective
  synchronization},}\ }\href {\doibase 10.1103/PhysRevLett.125.248101}
  {\bibfield  {journal} {\bibinfo  {journal} {Phys. Rev. Lett.}\ }\textbf
  {\bibinfo {volume} {125}},\ \bibinfo {pages} {248101} (\bibinfo {year}
  {2020})}\BibitemShut {NoStop}%
\bibitem [{\citenamefont {{Gast}}\ \emph {et~al.}(2020)\citenamefont {{Gast}},
  \citenamefont {{Schmidt}},\ and\ \citenamefont {{Knoesche}}}]{GSK20}%
  \BibitemOpen
  \bibfield  {author} {\bibinfo {author} {\bibfnamefont {R.}~\bibnamefont
  {{Gast}}}, \bibinfo {author} {\bibfnamefont {H.}~\bibnamefont {{Schmidt}}}, \
  and\ \bibinfo {author} {\bibfnamefont {T.~R.}\ \bibnamefont {{Knoesche}}},\
  }\bibfield  {title} {\enquote {\bibinfo {title} {A mean-field description of
  bursting dynamics in spiking neural networks with short-term adaptation},}\
  }\href@noop {} {\bibfield  {journal} {\bibinfo  {journal} {Neural
  Computation}\ }\textbf {\bibinfo {volume} {32}},\ \bibinfo {pages}
  {1615--1634} (\bibinfo {year} {2020})}\BibitemShut {NoStop}%
\bibitem [{\citenamefont {Goldobin}\ \emph {et~al.}(2021)\citenamefont
  {Goldobin}, \citenamefont {di~Volo},\ and\ \citenamefont
  {Torcini}}]{goldobin21}%
  \BibitemOpen
  \bibfield  {author} {\bibinfo {author} {\bibfnamefont {D.~S.}\ \bibnamefont
  {Goldobin}}, \bibinfo {author} {\bibfnamefont {M.}~\bibnamefont {di~Volo}}, \
  and\ \bibinfo {author} {\bibfnamefont {A.}~\bibnamefont {Torcini}},\
  }\bibfield  {title} {\enquote {\bibinfo {title} {Reduction methodology for
  fluctuation driven population dynamics},}\ }\href {\doibase
  10.1103/PhysRevLett.127.038301} {\bibfield  {journal} {\bibinfo  {journal}
  {Phys. Rev. Lett.}\ }\textbf {\bibinfo {volume} {127}},\ \bibinfo {pages}
  {038301} (\bibinfo {year} {2021})}\BibitemShut {NoStop}%
\bibitem [{\citenamefont {Chen}\ and\ \citenamefont {Campbell}(2022)}]{chen22}%
  \BibitemOpen
  \bibfield  {author} {\bibinfo {author} {\bibfnamefont {L.}~\bibnamefont
  {Chen}}\ and\ \bibinfo {author} {\bibfnamefont {S.~A.}\ \bibnamefont
  {Campbell}},\ }\bibfield  {title} {\enquote {\bibinfo {title} {Exact
  mean-field models for spiking neural networks with adaptation},}\ }\href@noop
  {} {\bibfield  {journal} {\bibinfo  {journal} {Journal of Computational
  Neuroscience}\ }\textbf {\bibinfo {volume} {50}},\ \bibinfo {pages}
  {445--469} (\bibinfo {year} {2022})}\BibitemShut {NoStop}%
\bibitem [{\citenamefont {Pietras}\ \emph {et~al.}(2023)\citenamefont
  {Pietras}, \citenamefont {Cestnik},\ and\ \citenamefont
  {Pikovsky}}]{pietras23}%
  \BibitemOpen
  \bibfield  {author} {\bibinfo {author} {\bibfnamefont {B.}~\bibnamefont
  {Pietras}}, \bibinfo {author} {\bibfnamefont {R.}~\bibnamefont {Cestnik}}, \
  and\ \bibinfo {author} {\bibfnamefont {A.}~\bibnamefont {Pikovsky}},\
  }\bibfield  {title} {\enquote {\bibinfo {title} {Exact finite-dimensional
  description for networks of globally coupled spiking neurons},}\ }\href
  {\doibase 10.1103/PhysRevE.107.024315} {\bibfield  {journal} {\bibinfo
  {journal} {Phys. Rev. E}\ }\textbf {\bibinfo {volume} {107}},\ \bibinfo
  {pages} {024315} (\bibinfo {year} {2023})}\BibitemShut {NoStop}%
\bibitem [{\citenamefont {Clusella}\ and\ \citenamefont
  {Montbri\'o}(2024)}]{clusella24}%
  \BibitemOpen
  \bibfield  {author} {\bibinfo {author} {\bibfnamefont {P.}~\bibnamefont
  {Clusella}}\ and\ \bibinfo {author} {\bibfnamefont {E.}~\bibnamefont
  {Montbri\'o}},\ }\bibfield  {title} {\enquote {\bibinfo {title} {Exact
  low-dimensional description for fast neural oscillations with low firing
  rates},}\ }\href {\doibase 10.1103/PhysRevE.109.014229} {\bibfield  {journal}
  {\bibinfo  {journal} {Phys. Rev. E}\ }\textbf {\bibinfo {volume} {109}},\
  \bibinfo {pages} {014229} (\bibinfo {year} {2024})}\BibitemShut {NoStop}%
\bibitem [{\citenamefont {Coombes}\ and\ \citenamefont {Byrne}(2019)}]{CB19}%
  \BibitemOpen
  \bibfield  {author} {\bibinfo {author} {\bibfnamefont {S.}~\bibnamefont
  {Coombes}}\ and\ \bibinfo {author} {\bibfnamefont {{\'A}.}~\bibnamefont
  {Byrne}},\ }\bibfield  {title} {\enquote {\bibinfo {title} {Next generation
  neural mass models},}\ }in\ \href@noop {} {\emph {\bibinfo {booktitle}
  {Nonlinear Dynamics in Computational Neuroscience}}},\ \bibinfo {editor}
  {edited by\ \bibinfo {editor} {\bibfnamefont {F.}~\bibnamefont {Corinto}}\
  and\ \bibinfo {editor} {\bibfnamefont {A.}~\bibnamefont {Torcini}}}\
  (\bibinfo  {publisher} {Springer International Publishing},\ \bibinfo
  {address} {Cham},\ \bibinfo {year} {2019})\ pp.\ \bibinfo {pages}
  {1--16}\BibitemShut {NoStop}%
\bibitem [{\citenamefont {Coombes}\ and\ \citenamefont
  {Wedgwood}(2023)}]{Coombes}%
  \BibitemOpen
  \bibfield  {author} {\bibinfo {author} {\bibfnamefont {S.}~\bibnamefont
  {Coombes}}\ and\ \bibinfo {author} {\bibfnamefont {K.~C.~A.}\ \bibnamefont
  {Wedgwood}},\ }\href@noop {} {\emph {\bibinfo {title} {Neurodynamics: An
  Applied Mathematics Perspective}}},\ Vol.~\bibinfo {volume} {75}\ (\bibinfo
  {publisher} {Springer Nature},\ \bibinfo {year} {2023})\BibitemShut {NoStop}%
\bibitem [{\citenamefont {Rabuffo}\ \emph {et~al.}(2021)\citenamefont
  {Rabuffo}, \citenamefont {Fousek}, \citenamefont {Bernard},\ and\
  \citenamefont {Jirsa}}]{rabuffo21}%
  \BibitemOpen
  \bibfield  {author} {\bibinfo {author} {\bibfnamefont {G.}~\bibnamefont
  {Rabuffo}}, \bibinfo {author} {\bibfnamefont {J.}~\bibnamefont {Fousek}},
  \bibinfo {author} {\bibfnamefont {C.}~\bibnamefont {Bernard}}, \ and\
  \bibinfo {author} {\bibfnamefont {V.}~\bibnamefont {Jirsa}},\ }\bibfield
  {title} {\enquote {\bibinfo {title} {Neuronal cascades shape whole-brain
  functional dynamics at rest},}\ }\href {\doibase 10.1523/ENEURO.0283-21.2021}
  {\bibfield  {journal} {\bibinfo  {journal} {eNeuro}\ }\textbf {\bibinfo
  {volume} {8}} (\bibinfo {year} {2021}),\
  10.1523/ENEURO.0283-21.2021}\BibitemShut {NoStop}%
\bibitem [{\citenamefont {Gerster}\ \emph {et~al.}(2021)\citenamefont
  {Gerster}, \citenamefont {Taher}, \citenamefont {{\v{S}}koch}, \citenamefont
  {Hlinka}, \citenamefont {Guye}, \citenamefont {Bartolomei}, \citenamefont
  {Jirsa}, \citenamefont {Zakharova},\ and\ \citenamefont {Olmi}}]{gerster22}%
  \BibitemOpen
  \bibfield  {author} {\bibinfo {author} {\bibfnamefont {M.}~\bibnamefont
  {Gerster}}, \bibinfo {author} {\bibfnamefont {H.}~\bibnamefont {Taher}},
  \bibinfo {author} {\bibfnamefont {A.}~\bibnamefont {{\v{S}}koch}}, \bibinfo
  {author} {\bibfnamefont {J.}~\bibnamefont {Hlinka}}, \bibinfo {author}
  {\bibfnamefont {M.}~\bibnamefont {Guye}}, \bibinfo {author} {\bibfnamefont
  {F.}~\bibnamefont {Bartolomei}}, \bibinfo {author} {\bibfnamefont
  {V.}~\bibnamefont {Jirsa}}, \bibinfo {author} {\bibfnamefont
  {A.}~\bibnamefont {Zakharova}}, \ and\ \bibinfo {author} {\bibfnamefont
  {S.}~\bibnamefont {Olmi}},\ }\bibfield  {title} {\enquote {\bibinfo {title}
  {Patient-specific network connectivity combined with a next generation neural
  mass model to test clinical hypothesis of seizure propagation},}\ }\href@noop
  {} {\bibfield  {journal} {\bibinfo  {journal} {Frontiers in Systems
  Neuroscience}\ }\textbf {\bibinfo {volume} {15}},\ \bibinfo {pages} {675272}
  (\bibinfo {year} {2021})}\BibitemShut {NoStop}%
\bibitem [{\citenamefont {Chandra}\ \emph {et~al.}(2019)\citenamefont
  {Chandra}, \citenamefont {Girvan},\ and\ \citenamefont {Ott}}]{chandra19}%
  \BibitemOpen
  \bibfield  {author} {\bibinfo {author} {\bibfnamefont {S.}~\bibnamefont
  {Chandra}}, \bibinfo {author} {\bibfnamefont {M.}~\bibnamefont {Girvan}}, \
  and\ \bibinfo {author} {\bibfnamefont {E.}~\bibnamefont {Ott}},\ }\bibfield
  {title} {\enquote {\bibinfo {title} {{Complexity reduction ansatz for systems
  of interacting orientable agents: Beyond the Kuramoto model}},}\ }\href
  {\doibase 10.1063/1.5093038} {\bibfield  {journal} {\bibinfo  {journal}
  {Chaos}\ }\textbf {\bibinfo {volume} {29}},\ \bibinfo {pages} {053107}
  (\bibinfo {year} {2019})}\BibitemShut {NoStop}%
\bibitem [{\citenamefont {Watanabe}\ and\ \citenamefont
  {Strogatz}(1993)}]{WS93}%
  \BibitemOpen
  \bibfield  {author} {\bibinfo {author} {\bibfnamefont {S.}~\bibnamefont
  {Watanabe}}\ and\ \bibinfo {author} {\bibfnamefont {S.~H.}\ \bibnamefont
  {Strogatz}},\ }\bibfield  {title} {\enquote {\bibinfo {title} {Integrability
  of a globally coupled oscillator array},}\ }\href {\doibase
  10.1103/PhysRevLett.70.2391} {\bibfield  {journal} {\bibinfo  {journal}
  {Phys. Rev. Lett.}\ }\textbf {\bibinfo {volume} {70}},\ \bibinfo {pages}
  {2391--2394} (\bibinfo {year} {1993})}\BibitemShut {NoStop}%
\bibitem [{\citenamefont {Cestnik}\ and\ \citenamefont
  {Martens}(2024)}]{rok24}%
  \BibitemOpen
  \bibfield  {author} {\bibinfo {author} {\bibfnamefont {R.}~\bibnamefont
  {Cestnik}}\ and\ \bibinfo {author} {\bibfnamefont {E.~A.}\ \bibnamefont
  {Martens}},\ }\bibfield  {title} {\enquote {\bibinfo {title} {Integrability
  of a globally coupled complex {R}iccati array: Quadratic integrate-and-fire
  neurons, phase oscillators, and all in between},}\ }\href {\doibase
  10.1103/PhysRevLett.132.057201} {\bibfield  {journal} {\bibinfo  {journal}
  {Phys. Rev. Lett.}\ }\textbf {\bibinfo {volume} {132}},\ \bibinfo {pages}
  {057201} (\bibinfo {year} {2024})}\BibitemShut {NoStop}%
\bibitem [{\citenamefont {Paz{\'o}}(2024)}]{pazo24}%
  \BibitemOpen
  \bibfield  {author} {\bibinfo {author} {\bibfnamefont {D.}~\bibnamefont
  {Paz{\'o}}},\ }\bibfield  {title} {\enquote {\bibinfo {title}
  {Quasi-integrable arrays: The family grows},}\ }\href {\doibase
  10.1103/Physics.17.12} {\bibfield  {journal} {\bibinfo  {journal} {Physics}\
  }\textbf {\bibinfo {volume} {17}},\ \bibinfo {pages} {12} (\bibinfo {year}
  {2024})}\BibitemShut {NoStop}%
\bibitem [{Note1()}]{Note1}%
  \BibitemOpen
  \bibinfo {note} {If the sign of $\Gamma +f_i-{b_r b_i}/(2a)$ changes under
  the time evolution, then our approach breaks down (one cannot simply flip the
  sign of the $\pm $). Fortunately, that does not occur in many situations. See
  Supplemental Material at xxx for details.}\BibitemShut {Stop}%
\bibitem [{\citenamefont {Ermentrout}\ and\ \citenamefont
  {Terman}(2010)}]{ET10}%
  \BibitemOpen
  \bibfield  {author} {\bibinfo {author} {\bibfnamefont {G.~B.}\ \bibnamefont
  {Ermentrout}}\ and\ \bibinfo {author} {\bibfnamefont {D.~H.}\ \bibnamefont
  {Terman}},\ }\href@noop {} {\emph {\bibinfo {title} {Mathematical foundations
  of neuroscience}}},\ Vol.~\bibinfo {volume} {64}\ (\bibinfo  {publisher}
  {Springer},\ \bibinfo {year} {2010})\BibitemShut {NoStop}%
\bibitem [{\citenamefont {Pikovsky}\ and\ \citenamefont
  {Politi}(2016)}]{PP_LEs}%
  \BibitemOpen
  \bibfield  {author} {\bibinfo {author} {\bibfnamefont {A.}~\bibnamefont
  {Pikovsky}}\ and\ \bibinfo {author} {\bibfnamefont {A.}~\bibnamefont
  {Politi}},\ }\href@noop {} {\emph {\bibinfo {title} {Lyapunov Exponents}}}\
  (\bibinfo  {publisher} {Cambridge University Press},\ \bibinfo {address}
  {Cambridge, UK},\ \bibinfo {year} {2016})\BibitemShut {NoStop}%
\bibitem [{Note2()}]{Note2}%
  \BibitemOpen
  \bibinfo {note} {This is precisely the marginal distribution for a
  conditional Lorentzian distribution $L(\eta |\eta _c,\Delta )$ centered at
  $\eta _c$ with HWHM $\Delta $, if $\eta _c$ is itself Lorentzian-distributed
  with center $\eta _0$ and HWHM $\delta $: $\DOTSI \intop \ilimits@ d\eta _c
  L(\eta |\eta _c,\Delta ) L(\eta _c|\eta _0,\delta ) =L(\eta |\eta _0,\delta
  +\Delta )$.}\BibitemShut {Stop}%
\bibitem [{Note3()}]{Note3}%
  \BibitemOpen
  \bibinfo {note} {The microscopic model of QIF neurons leading to Eq.~\protect
  \textup {\hbox {\mathsurround \z@ \protect \normalfont (\ignorespaces \ref
  {nmme}\unskip \@@italiccorr )}}, via \cite {pietras19,MP20}, is available in
  the Supplemental Material at XXXX.}\BibitemShut {Stop}%
\bibitem [{Note4()}]{Note4}%
  \BibitemOpen
  \bibinfo {note} {See Supplemental Material at XXX.}\BibitemShut {Stop}%
\bibitem [{\citenamefont {Th\"umler}\ \emph {et~al.}(2023)\citenamefont
  {Th\"umler}, \citenamefont {Srinivas}, \citenamefont {Schr\"oder},\ and\
  \citenamefont {Timme}}]{complexified2023}%
  \BibitemOpen
  \bibfield  {author} {\bibinfo {author} {\bibfnamefont {M.}~\bibnamefont
  {Th\"umler}}, \bibinfo {author} {\bibfnamefont {S.~G.~M.}\ \bibnamefont
  {Srinivas}}, \bibinfo {author} {\bibfnamefont {M.}~\bibnamefont
  {Schr\"oder}}, \ and\ \bibinfo {author} {\bibfnamefont {M.}~\bibnamefont
  {Timme}},\ }\bibfield  {title} {\enquote {\bibinfo {title} {Synchrony for
  weak coupling in the complexified {K}uramoto model},}\ }\href {\doibase
  10.1103/PhysRevLett.130.187201} {\bibfield  {journal} {\bibinfo  {journal}
  {Phys. Rev. Lett.}\ }\textbf {\bibinfo {volume} {130}},\ \bibinfo {pages}
  {187201} (\bibinfo {year} {2023})}\BibitemShut {NoStop}%
\bibitem [{\citenamefont {Lee}\ \emph {et~al.}(2024)\citenamefont {Lee},
  \citenamefont {Braun}, \citenamefont {Bönisch}, \citenamefont {Schröder},
  \citenamefont {Thümler},\ and\ \citenamefont {Timme}}]{complexified2024}%
  \BibitemOpen
  \bibfield  {author} {\bibinfo {author} {\bibfnamefont {S.}~\bibnamefont
  {Lee}}, \bibinfo {author} {\bibfnamefont {L.}~\bibnamefont {Braun}}, \bibinfo
  {author} {\bibfnamefont {F.}~\bibnamefont {Bönisch}}, \bibinfo {author}
  {\bibfnamefont {M.}~\bibnamefont {Schröder}}, \bibinfo {author}
  {\bibfnamefont {M.}~\bibnamefont {Thümler}}, \ and\ \bibinfo {author}
  {\bibfnamefont {M.}~\bibnamefont {Timme}},\ }\bibfield  {title} {\enquote
  {\bibinfo {title} {{Complexified synchrony}},}\ }\href {\doibase
  10.1063/5.0205897} {\bibfield  {journal} {\bibinfo  {journal} {Chaos}\
  }\textbf {\bibinfo {volume} {34}},\ \bibinfo {pages} {053141} (\bibinfo
  {year} {2024})}\BibitemShut {NoStop}%
\bibitem [{Note5()}]{Note5}%
  \BibitemOpen
  \bibinfo {note} {Reversible means that the system remains invariant under
  time inversion $t\to -t$ combined with one involution $\protect \cal G$ (a
  transformation which applied twice is equal to the identity: ${\protect \cal
  G} \circ {\protect \cal G} = 1$). The relevant involution $\protect \cal G$
  is $\beta \to -\beta -b_r/a$. Each neutral periodic orbits swirls around one
  of the two neutral foci located on the invariant subspace of the involution
  $\beta _f=-b_r/(2a)$--- provided $\eta _r>\eta _d=b_r^2/(4a)-f_r+a\alpha
  ^2$.}\BibitemShut {Stop}%
\end{thebibliography}


 \clearpage

 \begin{widetext}
\begin{center}
 {\large{Supplemental Material to \\
``Low Dimensional Dynamics of Globally Coupled Complex Riccati Equations: \\
Exact Firing-rate Equations for Spiking Neurons with Clustered Substructure''}}
\end{center}

\makeatletter
\setcounter{equation}{0}
\setcounter{figure}{0}
\renewcommand{\theequation}{S\arabic{equation}}
\renewcommand{\thefigure}{S\@arabic\c@figure}
\makeatother

 \setcounter{tocdepth}{1}
% \tableofcontents

\section{Simulation details}

\subsection{Initializing units with prescribed density}

We desire initializing units spread according to the density:
\begin{equation}
\rho(z)=\frac{1/\pi}{(|z|^2+1)^2}=\frac{1/\pi}{(1+x^2+y^2)^2}
\label{density}
\end{equation}
We restrict to $q=0$ and $\alpha=1$, since the general case just requires a trivial shift and rescaling.

\subsubsection{Relation with the uniform density on a sphere}

We first note that this density is obtained from a uniform density on the sphere $\tilde\rho(\theta,\phi)=(4\pi)^{-1}$
under stereographic projection:
\begin{equation}
x=\frac{\sin\theta \cos\phi}{1-\cos\theta} \qquad ; \qquad y=\frac{\sin\theta \sin\phi}{1-\cos\theta}
\label{stereo}
\end{equation}
Variables $\theta$ and $\phi$ are the polar and azimuthal angles, respectively.
We adopt the variable $c\equiv \cos\theta$, such that the differential element of surface $\sin\theta \, d\theta \, d\phi$
becomes $dc \, d\phi$, while $\tilde\rho(c,\phi)=(4\pi)^{-1}$.
The sought density is found after trivial but convoluted operations following the standard transformation:
\begin{equation}
\rho(x,y)=\tilde\rho(c.\phi) \left|\frac{\partial(c,\phi)}{\partial(x,y)} \right|
\end{equation}

% \subsection{Initialization}

As mentioned, the density \eqref{density} is the result of projecting a uniform density on the sphere.
Random drawing from the distribution \eqref{density} is performed taking $\theta$ and $\phi$ at random
and apply the transformation in \eqref{stereo}.
While $\phi$ is uniformly sampled in the interval $[0,2\pi)$,
some care must be taken with $\theta$. The differential $dc\equiv d\cos\theta$ dictates
we have to take a random quantity $w$ in the interval $[-1,1]$ and obtain the polar angle as $\theta=\arccos w$.

\subsubsection{Initialization (alternative)}

Working in polar coordinates $z=r \, e^{i\varphi}$, we can take $\varphi$ at random. For the radius
note that the radial density reads
\begin{equation}
p(r)=\frac{2}{(1+r^2)^2}
\end{equation}
The cumulative distribution function is
\begin{equation}
P(r)=\int_0^r p(r') \, r'\, dr'=\frac{r^2}{1+r^2}
\end{equation}
Sampling a uniform random variable $w$ in the unit interval, we get $r=\sqrt{w/(1-w)}$.

\subsection{Sampling of $\eta_j$}

The values of the $\{\eta_j\}_{j=1,\ldots,N}$ were selected deterministically
using quantiles of the Lorentzian distribution:

\begin{equation}
 \eta_j=\eta_0+\delta \tan[(\pi/2)(2j-N-1)/(N+1)]
\end{equation}

\subsection{Numerical integration}

For the simulation of the ensemble of Riccati equations we used a Runge-Kutta integrator
with adaptive stepsize control. The nominal stepsize was in the range $10^{-3}-10^{-4}$.

% \newpage

\section{Calculation of the mean field}

We calculate the mean field replacing the integral along the real axis
by a contour integration $\oint$ defined closing the real line by an arc at infinity
(either in the upper or in the lower half-plane $\eta=|\eta|e^{i\theta}\in\mathbb{C}$):
\begin{equation}
Z = \int_{-\infty}^\infty g(\eta) q(\eta,t) \, d\eta=\oint g(\eta) \, q(\eta,t) \, d\eta  - \lim_{|\eta|\to\infty} i |\eta|
\int_0^{\pm\pi} g(\eta) \, q(\eta,t) \,e^{i\theta} d\theta = q(\eta_p,t)
\label{residue}
\end{equation}
The last equality is true if:
\begin{enumerate}
 \item The only pole ($\eta_p=\eta_0\pm i \delta$) inside the integration contour
 comes from the Lorentzian distribution
\begin{equation}
g(\eta)=\frac{\delta/\pi}{(\eta-\eta_0)^2+
\delta^2}=\frac1{2\pi i}\left[\frac1{\eta-\eta_0-i\delta}-\frac1{\eta-\eta_0+i\delta}\right] .
\label{lorr}
\end{equation}
We analyze below in which half-plane is $q$ analytic, i.e.~if $q(\eta,0)$ is initially analytic it remains so.

\item  The integration along the arc vanishes.
\end{enumerate}

\subsection{Analytic continuation}

The application of the residue theorem requires $q$ to possess an analytic continuation
from the real $\eta$ to
complex $\eta=\eta_r+i\eta_i$ in one half-plane  ($\eta_i>0$ or $\eta_i<0$).

We write the equation for $q=\beta+i\gamma$ in real variable:
\begin{subequations}
\begin{eqnarray}
\partial_t\beta&=& a\left (\beta^2-\gamma^2\right)+  b_r \, \beta - b_i\gamma
+ \eta_r + f_r- a\alpha^2 \\
\partial_t\gamma&=& 2 a\,\beta \gamma
+ b_r \gamma + b_i\beta  + \eta_i + \Gamma + f_i
\end{eqnarray}
\label{2eq}
\end{subequations}
It is not obvious at first sight how the time evolution of $q$, governed by
these equations, could break the analyticity of a certain
analytic initial condition $q(\eta,0)$. We gain insight assuming constant
$\alpha$ and $f$.
The key observation is that
at
\begin{equation}
\eta_i=-\Gamma-f_i + \frac{b_r b_i}{2a} \label{cond}
\end{equation}
the system in Eq.~\eqref{2eq} is reversible. This is relevant
on the half-line  $\eta_r\in(\eta_d,\infty)$, since it
exhibits a continuum of periodic orbits\footnote{Reversible
means that the system remains invariant under time inversion $t\to -t$ combined with
one involution $\cal G$ (a transformation which applied twice is
equal to the identity: ${\cal G} \circ {\cal G} = 1$).
The relevant involution $\cal G$ is  $\beta\to -\beta-b_r/a$. Each neutral periodic
orbits swirls around one of the two neutral foci
located on the invariant subspace of the involution $\beta_f=-b_r/(2a)$---
provided $\eta_r>\eta_d=b_r^2/(4a)-f_r+a\alpha^2$.}.
In turn, there is a discontinuity in $q(\eta,t)$
when crossing this half-line \eqref{cond}
in the complex $\eta$-plane.
This is something we avoid choosing the opposite half-plane when applying the residue theorem in Eq.~\eqref{residue}.

\subsection{Vanishing integral along the arc}

We analyze next the behavior of $q$ in the large $|\eta|$ limit.
The governing equations become:
\begin{subequations}
\begin{eqnarray}
\partial_t\beta&=& \beta^2-\gamma^2 +
|\eta| \cos\theta \\
\partial_t\gamma&=&  2 \beta \, \gamma + |\eta|\sin\theta
\end{eqnarray}
\end{subequations}
As we wish to know the behavior of the system at $|\eta|\to\infty$ the
linear and finite terms have been neglected.
It is easy to ascertain that there is a stable fixed point at
$(\beta_*,\gamma_*)=(\tilde\beta,\tilde\gamma)|\eta|^{1/2}$,
with $\tilde\beta,\tilde\gamma\sim O(1)$ and $\tilde\beta<0$.
This immediately implies that the integration along the arc in Eq.~\eqref{residue}  vanishes since
$g(\eta)\sim |\eta|^{-2}$\\

\section{Derivation of the ensemble of firing-rate equations from spiking neurons}

The microscopic model consists of $N$ subpopulations of $M$ QIF neurons each. Each neuron is identified by two indices $i$
and $j$, indicating  the neuron number
and population number, respectively.
A neuron receives inputs from its own subpopulation
(through the firing rate $r_j$), as
well as a global input from all neurons.

\subsection{Global chemical coupling}

We start with the system of QIF neurons in Eqs.~(13)-(15).
\\

The heterogeneity in each subpopulation is Lorentzian with half-width at half-maximum (HWHM)
$\Delta$. The center $\bar\eta_j$ is itself Lorentzian distributed:
\begin{eqnarray}
h(\eta_{i,j}|\bar \eta_j)&=&\frac{\Delta/\pi}{(\eta_{i,j}-\bar\eta_j)^2+\Delta^2} \label{h}\\
g(\bar \eta_j)&=&\frac{\delta/\pi}{(\bar\eta_{j}-\eta_0)^2+\delta^2}
\end{eqnarray}

In the thermodynamic limit, the system of equations (13), (14), \eqref{h} reduces,
via the theory developed in \cite{MPR15}, to the
ensemble of firing-rate models, Eq.~(16) in the main text. There, $\bar\eta_j$
has been replaced by $\eta_j$ to lighten the notation.\\

Note that the distribution of $\eta_{i,j}$ among all the population is a Lorentzian
centered at $\eta_0$ with  HWHM  $\Delta+\delta$:
$$
\int_{-\infty}^{\infty} h(\eta|\bar\eta) g(\bar\eta) d\bar\eta=
\frac{(\Delta+\delta)/\pi}{(\eta-\eta_0)^2+(\Delta+\delta)^2}
$$

\subsection{Global electrical coupling}

The system of equations is:
\begin{eqnarray}
{\dot V}_{i,j}&=&V_{i,j}^2+\eta_{i,j} + \phi(r_j) + g(V-V_{i,j}) \label{ve}\\
v_j&=&\frac{1} M \sum_{i=1}^M V_{i,j}\label{vje}\\
V&=&\frac{1}{N}  \sum_{j=1}^N v_{j}
\end{eqnarray}
Using the theory developed in \cite{MPR15,pietras19}
with symmetric spikes \cite{MP20} we get:
\begin{subequations}
 \begin{eqnarray}
\dot v_j&=&v_j^2 -\pi^2 r_j^2   + \phi(r_j) + \eta_j  + g (V-v_j) \\
\dot r_j&=& 2   v_j r_j+ \frac{\Delta}\pi -g r_j
\end{eqnarray}
\label{nmmee}
\end{subequations}

For $\phi(r)=\kappa r^2$, and the transformation
$z_j=v_j+i(\pi^2-\kappa)^{1/2}r_j$,
the ensemble of Riccati equations becomes:
\begin{equation}
% \dot z_j=z_j^2-g_j z_j +\eta_j + g v_{syn}+ i\Delta + J \, R
\dot z_j=z_j^2-g z_j+\eta_j + i(1-\kappa/\pi^2)^{1/2} \Delta + g V
\label{nmmce}
\end{equation}
With our reduction method we get:
\begin{equation}
\dot Z=Z^2-g\, Z+\eta_0+ i\left[(1-\kappa/\pi^2)^{1/2} \Delta + \delta\right]+ g V
\end{equation}
Recalling that $Z=V+i(\pi^2-\kappa)^{1/2}R$ we
get the macroscopic rate equations (21) in the main text.

\section{Several layers of networks and heterogeneity}
In the main text,
we study an ensemble of ensembles of QIF neurons. This
effectively creates a two-layer system ---i.e., a network of networks.
This approach can be extended to systems with multiple scales.
In this section, we present the formulation
for a three-layer system: micro, meso, and macro.\\

At the micro level, we consider individual QIF neurons, indexed by three variables: $i$, $j$, and $k$, representing the micro, meso, and macro levels, respectively. Thus, variable $V_{ijk}$ denotes the voltage of a neuron in the macro population ``$k$'', within the meso population ``$j$'', and with the microscopic index ``$i$''. \\

The QIF equation for every neuron is given by:
\begin{equation}
\dot{V}_{ijk} = V_{ijk}^2 + \phi_0(r_{jk}) + \phi_1(r_k) + \phi_2(R) + \eta_{ijk}
\label{eq:voltage}
\end{equation}
Here, $\phi_0$, $\phi_1$, and $\phi_2$ represent the coupling terms at the micro, meso, and macro scales, respectively. The term $r_{jk}$ represents the firing rate of the meso-macro population labeled by ``$jk$'', $r_k$ is the firing rate of the macro population ``$k$'', and $R$ denotes the global firing rate. These are defined as follows:
\begin{equation}
r_{jk} = \frac{1}{N_0}\sum\limits_{i=1}^{N_0} \sum\limits_{\ell} \frac{1}{\tau} \int\limits_{t-\tau}^t \delta(s-t_{ijk}^\ell) ds, \qquad r_{k} = \frac{1}{N_1} \sum\limits_{j=1}^{N_1} r_{jk}, \qquad R = \frac{1}{N_2} \sum\limits_{k=1}^{N_2} r_k
\end{equation}
where $t_{ijk}^\ell$ is the time of the $\ell^\text{th}$ spike of neuron ``$ijk$'', and we consider the infinite limit $N_0,N_1,N_2 \to \infty$ and $\tau \to 0$.
Similarly, the mean voltages at each level are expressed as:
\begin{equation}
v_{jk} = \frac{1}{N_0}\sum\limits_{i=1}^{N_0} V_{ijk}, \qquad v_k = \frac{1}{N_1}\sum\limits_{j=1}^{N_1} v_{jk}, \qquad V = \frac{1}{N_2}\sum\limits_{k=1}^{N_2} v_{k}
\end{equation}
The heterogeneity $\eta_{ijk} \in \mathbb{R}$ follows a Lorentzian distribution at all levels:
\begin{equation}
g_0(\eta_{ijk}) = \frac{\Delta_0/\pi}{(\eta_{ijk}-\eta_{jk})^2 + \Delta_0^2}, \quad g_1(\eta_{jk}) = \frac{\Delta_1/\pi}{(\eta_{jk}-\eta_k)^2 + \Delta_1^2}, \quad g_2(\eta_k) = \frac{\Delta_2/\pi}{(\eta_k - \eta)^2 + \Delta_2^2}
\end{equation}
where parameters $\Delta_0$, $\Delta_1$, and $\Delta_2$ control the width of the Lorentzian distributions at the micro, meso, and macro scales, respectively.

\subsection{Layer 1: Meso-Scale Dynamics via Lorentzian Ansatz}
We now apply the Lorentzian ansatz formalism~\cite{MPR15} to the voltage equation~\eqref{eq:voltage} to describe the dynamics at the meso scale. The resulting equation is:
\begin{equation}
\dot{z}_{jk} = z_{jk}^2 + \phi_0 + \phi_1 + \phi_2 + \eta_{jk} + i\Delta_0
\end{equation}
Here, the complex variable $z_{jk}$ represents both the mean voltage and firing rate, with $z_{jk} = v_{jk}+i\pi r_{jk}$. By expressing the dynamics in terms of real quantities, we obtain:
\begin{subequations}
\begin{align}
\dot{v}_{jk} &= v_{jk}^2 - \pi^2 r_{jk}^2 + \phi_0 + \phi_1 + \phi_2 + \eta_{jk}\\
\dot{r}_{jk} &= 2v_{jk} r_{jk} + \frac{\Delta_0}{\pi}
\end{align}
\end{subequations}
Now we choose the coupling at the micro level to be $\phi_0(r_{jk}) = \kappa_0 r_{jk}^2$ and change the dynamic variables with a transformation:
\begin{equation}
q_{jk} = v_{jk} + ir_{jk}\sqrt{\pi^2-\kappa_0}
\label{eq:q_trans}
\end{equation}
Under this transformation, we obtain a new system of coupled Riccati equations for the variables $q_{jk}$:
\begin{equation}
\dot{q}_{jk} = q_{jk}^2+\phi_1+\phi_2 + i\Delta_0\sqrt{1-\kappa_0/\pi} + \eta_{jk}
\label{eq:eqq}
\end{equation}

\subsection{Layer 2: Macro-Scale Dynamics} %via Extended Lorentzian Ansatz
Now that the Riccati array~\eqref{eq:eqq} is complex, we employ the formalism detailed in the main text. Moving to the next scale, we describe the dynamics as follows:
\begin{equation}
\dot{q}_{k} = q_{k}^2+\phi_1+\phi_2 + i\Delta_0\sqrt{1-\kappa_0/\pi} + i\Delta_1 \eta_{k}
\end{equation}
It is important to note that $q_k$ is not simply the complex voltage variable for this scale, i.e.~$q_k \neq v_k + i\pi r_k$. Instead we have to consider the inverse transformation~\eqref{eq:q_trans}, and write the mean voltage and firing rate equation at the macro scale:
\begin{subequations}
\begin{align}
\dot{v}_{k} &= v_{k}^2 - (\pi^2-\kappa_0)r_{k}^2 + \phi_1 + \phi_2 + \eta_{k}\\
\dot{r}_{k} &= 2v_{k} r_{k} + \frac{\Delta_0}{\pi} + \frac{\Delta_1}{\sqrt{\pi^2-\kappa_0}}
\end{align}
\end{subequations}
These equations are analogous to those derived in the main text, see Eqs.~(19).

Now in order to get a Riccati array again, we choose the coupling at the meso level to be $\phi_1(r_{k}) = \kappa_1 r_{k}^2$ and change the dynamic variables with another transformation:
\begin{equation}
w_{k} = v_{k} + ir_k \sqrt{\pi^2-\kappa_0-\kappa_1}
\label{eq:q_trans2}
\end{equation}
and obtain yet another system of coupled Riccati equations for the variables $w_{k}$:
\begin{equation}
\dot{w}_{k} = w_{k}^2+\phi_2 + i\Delta_0\sqrt{1-\kappa_0/\pi^2-\kappa_1/\pi^2} + i \Delta_1\sqrt{1-\frac{\kappa_1}{\pi^2-\kappa_0}}+ \eta_{k}
\label{eq:eqw}
\end{equation}

\subsection{Layer 3: Global Dynamics} % via (another) Extended Lorentzian Ansatz}
Following the approach from the previous subsection, we again apply the our ansatz to describe the dynamics at the global level. The evolution of the global variable is given by:
\begin{equation}
\dot{w} = w^2+\phi_2 + i\Delta_0\sqrt{1-\kappa_0/\pi^2-\kappa_1/\pi^2} + i \Delta_1\sqrt{1-\frac{\kappa_1}{\pi^2-\kappa_0}}+ i \Delta_2 + \eta
\end{equation}
which relates relates to the global mean voltage $V$ and the global firing rate $R$ via the expression: $w = V+iR\sqrt{\pi^2-\kappa_0-\kappa_1}$.
The corresponding dynamics for $v$ and $r$ are:
\begin{subequations}
\begin{align}
\dot{V} &= V^2 - (\pi^2-\kappa_0-\kappa_1) R^2 + \phi_2 + \eta\\
\dot{R} &= 2V R + \frac{\Delta_0}{\pi} + \frac{\Delta_1}{\sqrt{\pi^2-\kappa_0}} + \frac{\Delta_2}{\sqrt{\pi^2-\kappa_0-\kappa_1}}
\end{align}
\end{subequations}
These equations describe the global behavior of the system in terms of the firing rate and mean voltage, incorporating the couplings and heterogeneities across all three scales.

\subsection{Generalization to arbitrary number of scales}
At this point the pattern is visible: for $M+1$ layers, the global firing rate and voltage equations are:
\begin{subequations}
\begin{align}
\dot{V} &= V^2 - (\pi^2-\kappa_0-\kappa_1- ...-\kappa_{M-1}) r^2 + \phi_M + \eta\\
\dot{R} &= 2V R + \frac{\Delta_0}{\pi} + \frac{\Delta_1}{\sqrt{\pi^2-\kappa_0}} + \frac{\Delta_2}{\sqrt{\pi^2-\kappa_0-\kappa_1}} + ... + \frac{\Delta_M}{\sqrt{\pi^2-\kappa_0-\kappa_1-...-\kappa_{M-1}}}
\end{align}
\end{subequations}

\end{widetext}

\end{document}